# An Adaptive feature mode decomposition based on a novel health indicator for bearing fault diagnosis

**Sumika Chauhan[1], Govind Vashishtha [1*], Rajesh Kumar[2], Radoslaw Zimroz[1] and Pradeep Kundu[3]**

[1]Faculty of Geoengineering, Mining and Geology, Wroclaw University of Science and Technology, Na Grobli 15, 50-421 Wroclaw, Poland

[2]Precision Metrology Laboratory, Department of Mechanical Engineering, Sant Longowal Institute of Engineering andTechnology, Longowal 148 106, India

[3]Department of Mechanical Engineering, KU Leuven, Campus Bruges, 8200 Bruges, Belgium

* Corresponding author, Email address: govindyudivashishtha@gmail.com

**Abstract**

The vibration analysis of the bearing is very crucial because of its non-stationary nature and low signal-to-noise ratio. Therefore, a novel scheme for detecting bearing defects is put forward based on the extraction of single-valued neutrosophic cross-entropy (SVNCE) to address this issue. Initially, the artificial hummingbird algorithm (AHA) is used to make the feature mode decomposition (FMD) adaptive by optimizing its parameter based on a newly developed health indicator (HI) i.e. sparsity impact measure index ($SIMI$). This HI ensures full sparsity and impact properties simultaneously. The raw signals are disintegrated into different modes by adaptive FMD at optimal values of its parameters. The energy of these modes is calculated for different health conditions. The energy interval range has been decided based on energy eigen which are then transformed into single-valued neutrosophic sets (SVNSs) for unknown defect conditions. The minimum argument principle employs the least SVNCE values between SVNSs of testing samples (obtained from unknown bearing conditions) and SVNSs of training samples (obtained from known bearing conditions) to recognize the different defects in the bearing. It has been discovered that the suggested methodology is more adept at identifying the various bearing defects.

**Keywords:** Bearing, feature mode decomposition, sparsity impact measure index, artificial hummingbird, neutrosophic set, symmetric cross-entropy

## 1. Introduction

The rotating elements such as bearings, gears and shafts are widely used in the industries. Due to continuous operation in harsh working environments, they are subjected to failures. The unforeseen failures of these components affect the performance which may lead to economic and disastrous consequences [1][2][3]. Out of these components, the bearings are critical components used in various machines and industrial equipment to support rotating shafts and reduce friction. They play a crucial role in ensuring smooth and efficient operation. However, over time, bearings can suffer from wear and tear, leading to faults and failures. Fault diagnosis of bearings is a vital process aimed at detecting, identifying, and assessing these faults before they escalate into catastrophic failures [4–7].

Over time, several fault diagnosis approaches have evolved. Out of which, vibration analysis of bearings is a critical technique that is used to monitor their health condition. The development of the bandpass filter with the help of spectral kurtosis (SK), fast kurtogram, and minimum entropy deconvolution (MED) along with their improved versions is used to analyse the signals obtained from the bearings [8–10]. Due to harsh working environments, noisy conditions and long transmission lines, sensitive information may get embedded within the noisy signals. Therefore, preprocessing of these signals becomes important for satisfactory diagnostic accuracy which can be accomplished by different decompositions techniques. Empirical mode decomposition (EMD), ensemble empirical mode decomposition (EEMD), variational mode decomposition (VMD), and local mode decomposition (LMD) are some of the most popular decomposition techniques [11–14] These signal-processing techniques have made considerable progress in mechanical fault detection. For instance, Kumar et al. [15] proposed the non-parametric complimentary EEMD that eliminates the need to define the SNR of the noise and the number of ensembles while processing the signals. Dao et al. [16] hybridized the wavelet threshold and EEMD to process the acoustic signals of the hydro turbine under different flow conditions. Li et al. [17] proposed the bandwidth EMD and incorporated it with adaptive morphological analysis to diagnose the rolling bearings. Chauhan et al. [18] utilised complete EEMD to decompose the signals into an intrinsic mode function based on corrected conditional entropy in order to diagnose the bearing. Vashishtha and Kumar [19] have introduced the improved version of the time-varying filter EMD and hybridized it with kernel estimate for mutual information (*KEMI*) to diagnose the different defects of the Pelton wheel. Darong et al. [20] applied envelope demodulation to the signals obtained from local mode decomposition to detect the sensitive fault features. Besides the numerous advantages of these techniques, some drawbacks are still associated with these decomposition techniques. For instance, the mode mixing and end effect affect the performance of the EMD [21]. In improved versions of EMD, the above-mentioned limitations have only been reduced but not removed completely. Wavelet transform (WT) is also used to pre-process the signals through Fourier spectrum segmentation. Researchers have proposed improved versions of WT by modifying the spectral segmentation and optimal selection of the frequency band. The spectral segmentation is affected by environmental factors and the non-adaptive nature of the wavelet influences the performance of the WT [22]. VMD utilized the concept of the Wiener filter to decompose the raw signals into different modes based on bandwidth and center frequency. Researchers have suggested that VMD is better than WT, EMD and its improved version as it eliminates the issue of mode mixing and helps in extracting prominent fault features due to its non-recursive nature. Vashishtha and Kumar [23] have used the salp swarm algorithm (SSA) to optimize the VMD's parameter to identify the impeller defects in the centrifugal pump. Kumar et al. [24] introduced a dynamic degradation monitoring technique utilising VMD based on trigonometric entropy measure. But in the case of non-stationary vibration signals, the filter cut-off frequency of VMD is not time-varying which affects the performance of the VMD [21].

From the literature, it is clear that the famous signal decomposition techniques are associated with some drawbacks that affect their performance. These drawbacks have been addressed by the novel decomposition method named feature mode decomposition (FMD) introduced by Miao et al. [25]. FMD is based on the FIR filter bank and maximum correlation kurtosis deconvolution theory which makes it proficient to detect impulsiveness and periodicity concurrently. FMD works better than other decomposition techniques under noisy environments and the fault period is not obligatory during its operation. required and other interferences. FMD depends on the two parameters viz., mode number and filter length whose values should be set in advance. The improper selection of these two parameters affects the performance of FMD. Therefore, an optimization algorithm should be incorporated in the FMD to choose the mode number and filter length optimally. The main input of this research includes:

- An adaptive FMD is proposed to boost the adaptability corresponding to the input signals for refining the fault features extraction ability. The artificial hummingbird algorithm (AHA) has been used to serve this purpose.
- The fractional Gaussian noise (FGN) has been used to examine the filter properties of the FMD.
- A new health indicator (HI) named $SIMI$ (sparsity impact measure index) has been developed utilising approximate entropy ($ApEn$) and kurtosis index ($KI$) to guide the optimization process. $ApEn$ is sensitive to periodicity through which it directly measures the variation in the sparsity. Whereas, $KI$ reflects the variation of impact property. Thus, $SIMI$ analyses the fault characteristic information which characterizes the signal's periodic impulse as well as its degree of resistance to impulsive noises.

- The efficacy of the proposed methodology has been proved through the case study.

The remaining paper is designed in the following sections. The FMD is briefed in Section 2. Section 3 discusses how FMD is made adaptive through parameter optimization using AHA. This section also elaborates on the procedure of constructing the health indicator. Section 4 deals with the proposed methodology. In Section 5, the proposed methodology has been applied to real-world applications and the corresponding results have been discussed. Finally, the conclusions and highlights of the research work are given in Section 6.

## 2. Background

### 2.1. Feature mode decomposition

Miao et al. [25] have proposed the non-recursive decomposition techniques i.e. FMD that initializes the FIR filter bank by modifying the filter coefficients to select the number of modes adaptively. The FMD is implemented through the following steps.

**Step 1**: Input the raw signal $(x)$ into FMD at preset values of different parameters mode number $(K)$, filter length $(L)$ and maximum iteration $(Max_{iter})$.

**Step 2**: The FIR filter bank is initialized by Hanning windows having $M$ filters whose value is taken between 5 to 10 by setting the current iteration $(Iter)$ to 1.

**Step 3**: Use $u_m^i = x * f_m^i$ to disintegrate the signals into different modes, where $m = 1, 2, ..., M$ and $*$ is convolution operation.

**Step 4**: The filter coefficients should be updated according to the input signal $(x)$, mode components $u_m^i$ and faulty period $T_m^i$. $T_m^i$ is a time interval between the initial crossing and the local maximum of the auto-correlation spectrum $R_m^i$ of $u_m^i$.

**Step 5**: Check whether $Iter = Max_{iter}$. If this condition is met then execute Step 6. Otherwise, repeat the iterations from Step 3.

**Step 6**: The correlation coefficient $(CC)$ should be computed for every two modes to select the modes having the highest values of $CC$. Then correlation kurtosis of these modes is evaluated based on $T_m^i$. The mode with a larger value of correlation kurtosis is selected and set $M = M - 1$.

**Step 7**: Check whether the $M$ reaches the specified value of $K$. If this condition is met then execute Step 8. Otherwise, repeat the iterations from Step 3.

**Step 8:** Save the reserved modes as the final decomposed mode.

The filtering properties of the FMD is mainly depends on the values of mode number $K$ and filter length $L$ respectively. When $L$ is constant and $K$ is varying the corresponding filter bandwidth is found to be large and pass-band ripples are not encouraging to de-noising. On the other hand, when the $K$ is set to a constant value by changing the values of $L$, the pass-band ripples appear in this case. Through this discussion, it is suggested that the two parameters viz., $K$ and $L$ have a significant role in the decomposition ability of the FMD. Therefore, a prominent method should be adopted to select these parameters.

### 2.1. Symmetric single-valued neutrosophic cross-entropy (SVNCE) measure

The concept of SNVCE is taken from [26,27] and elaborated through subsequent theorems:

**Theorem 2.1**: The $T_N(A)$ is the SVNC measure having a minimum value of $T_N(A) = 0$ and the maximum value of $T_N(A) = 3\left(log_2 2 - log_2 \frac{5}{3}\right)n$,

$$T_N(A) = 2n\log_2 \frac{3}{5} + \sum_{i=1}^{n}\left[\log_2\left[1 + \frac{2}{5}\sqrt{i_A(x_i)(1 - i_A(x_i))}\right] + \frac{1}{3}(2 + \mu_A(x_i) + f_A(x_i)) \times \log_2\left[1 + \frac{2 + 2\sqrt{\mu_A(x_i)f_A(x_i)}}{2 + \mu_A(x_i) + f_A(x_i)}\right] + \frac{1}{3}(4 - \mu_A(x_i) - f_A(x_i)) \times \log_2\left[\frac{2 + 2\sqrt{(1-\mu_A(x_i))(1-f_A(x_i))}}{4 - \mu_A(x_i) - f_A(x_i)}\right]\right] \quad (1)$$

where the truth membership function is represented by $\mu_A(x_i)$, interdependency function is shown by $i_A(x_i)$, $f_A(x_i)$ indicates the falsity membership function, $x$ is the signal, and $n$ is number of datapoints in a signal.

**Theorem 2.2:** Consider $A, B \in W(X)$ are any two SVNSs, where $W(X)$ is a collection of SVNS the SVNCE measure $(T_{SVNCE}(A, B))$ is given as

$$T_{SVNCE}(A, B) = \sum_{i=1}^{n}\left[(2 + \mu_A(x_i) + \mu_B(x_i)) \times log_2\left[\frac{2 + \mu_A(x_i) + \mu_B(x_i)}{\frac{1}{2}[4 + \mu_A(x_i) + \mu_B(x_i) + 2\sqrt{\mu_A(x_i)\mu_B(x_i)}]}\right] + (4 - \mu_A(x_i) - \mu_B(x_i)) \times log_2\left[\frac{4 - \mu_A(x_i) - \mu_B(x_i)}{\frac{1}{2}[6 - \mu_A(x_i) - \mu_B(x_i) + 2\sqrt{(1-\mu_A(x_i))(1-\mu_B(x_i))}]}\right]\right] + \sum_{i=1}^{n}\left[(2 + i_A(x_i) + i_B(x_i)) \times log_2\left[\frac{2 + i_A(x_i) + i_B(x_i)}{\frac{1}{2}[4 + i_A(x_i) + i_B(x_i) + 2\sqrt{i_A(x_i)i_B(x_i)}]}\right] + (4 - i_A(x_i) - i_B(x_i)) \times log_2\left[\frac{4 - i_A(x_i) - i_B(x_i)}{\frac{1}{2}[6 - i_A(x_i) - i_B(x_i) + 2\sqrt{(1-i_A(x_i))(1-i_B(x_i))}]}\right]\right] + \sum_{i=1}^{n}\left[(2 + f_A(x_i) + f_B(x_i)) \times \right.$$

$$log_2\left[\frac{2+f_A(x_i)+f_B(x_i)}{\frac{1}{2}[4+f_A(x_i)+f_B(x_i)+2\sqrt{f_A(x_i)f_B(x_i)}]}\right] + (4-f_A(x_i)-f_B(x_i)) \times$$

$$log_2\left[\frac{4-f_A(x_i)-f_B(x_i)}{\frac{1}{2}[6-f_A(x_i)-f_B(x_i)+2\sqrt{(1-f_A(x_i))(1-f_B(x_i))}]}\right]\right] \tag{2}$$

**Theorem 2.3:** It is intriguing to observe that $T_{SVNCE}(A^C, B^C) = T_{SVNCE}(A, B)$ for each SVNS. Here, $C$ denotes the signal's concavity. The resulting **Theorem 2.1** produces

$$T_{SVNCE}(A, A^C) = 2\sum_{i=1}^{n}\left[18\left(log_2 2 - log_2 \frac{5}{3}\right) - 3\left(2 log_2 \frac{3}{5} + \frac{1}{3}(2 + \mu_A(x_i) + \tilde{f}_A(x_i)) \times \right.\right.$$

$$log_2\left[1 + \frac{2+2\sqrt{\mu_A(x_i)f_A(x_i)}}{2+\mu_A(x_i)+f_A(x_i)}\right] + \frac{1}{3}(4 - \mu_A(x_i) - f_A(x_i)) \times log_2\left[1 + \frac{2+\sqrt{(1-\mu_A(x_i))(1-f_A(x_i))}}{4+\mu_A(x_i)+f_A(x_i)}\right] +$$

$$log_2\left[1 + \frac{2}{5}\sqrt{i_A(x_i)(1-i_A(x_i))}\right]\right) \tag{3}$$

By hybridizing Eq. (1) and (3), the Eq. (4) is obtained:

$$T_{SVNCE}(A, A^C) = 18\left(log_2 2 - log_2 \frac{5}{3}\right)n - 6T_N(A)$$

$$\Rightarrow T_N(A) = 3\left(log_2 2 - log_2 \frac{5}{3}\right)n - \frac{1}{6}T_{SVNCE}(A, A^C) \geq 0$$

$$\Rightarrow 0 \leq T_{SVNCE}(A, A^C) \leq 18\left(log_2 2 - log_2 \frac{5}{3}\right)n \tag{4}$$

## 3. Procedure for making FMD adaptive

The goal of the proposed work is to make FMD adaptive to extract the sensitive features which are prone to defects by choosing the optimal parameters of FMD. This can be implemented by following the two criteria: 1) development of the health indicator that not only helps in detecting the faults at an early stage but also acts as a fitness function of the optimization algorithm and 2) utilising the appropriate optimization algorithm for optimal selection of FMD parameters.

### 3.1. Construction of Health Indicator

The HI is essential for making FMD adaptive to validate the efficacy of the decomposition results. In this section, the procedure of developing appropriate HI from approximate entropy ($ApEn$) and kurtosis index ($KI$) has been discussed. Pincus introduced the $ApEn$ to measure the dynamic characteristics of the complex and short signals at different scales and frequencies in a noisy environment without coarse-graining [28]. Consequently, it is viewed as a direct indicator of sparsity. The sparsity of the vibration signal increases due to periodicity and the corresponding value of $ApEn$ decreases. It is preferred over other entropies because it can easily identify the irregularities within the signal with less computational time. The $ApEn$ is the likelihood that a time series will produce new patterns as its dimensionality changes. For a given time series $x(n) = \{x(1), x(2), ..., x(N)\}$ of length $N$ having window size $m$ (also known as pattern length). Then, the $ApEn$ can be computed by Eq. (5).

$$ApEn = \lim_{N \to \infty}[\phi^m(r) - \phi^{m+1}(r)] \qquad (5)$$

where $m$ is the embedding dimension, $\phi$ represents the average frequency of all of the subsequence patterns in the sequence remaining close to each other and $r$ is the similarity of the $ApEn$. But $ApEn$ is associated with some limitations such as: 1) It typically ignores amplitude information, 2) Equalities, or equal values of the vibration signals, are ignored, and 3) Performance also be improved in harsh operating environments. Kurtosis index ($KI$) is utilised to measure the impulse intensity of the vibration signal. With vibration signal $x = \{x(1), x(2), ..., x(n)\}$, the $KI$ can be evaluated through Eq. (6).

$$KI = \frac{\frac{1}{N}\sum_{n=0}^{N} x^4(n)}{\left(\frac{1}{N}\sum_{n=0}^{N-1} x^2(n)\right)^2} \qquad (6)$$

$KI$ is better able to detect impact strengths. But if the vibration signal is embedded with both fault excited periodic impulse and noise impulse then $KI$ is affected by noisy impulse leading to incorrect evaluation of the defects [23]. Therefore, a novel HI has been designed to address the issues of $ApEn$ and $KI$ and to fully guarantee the optimal sparsity and impact properties of the obtained modes simultaneously. The developed HI can be evaluated as follows:

$$SIMI = \frac{ApEn}{KI} \qquad (7)$$

It can be noticed from Eq. (7) that the smaller value of the $SIMI$ depicts better sparsity and impact properties. The developed HI also signifies high-quality components with accurate decomposition results. Therefore developed HI is not used as a fitness function of the optimization algorithm but also utilised as the criteria for selecting sensitive modes.

## 3.2. Artificial Hummingbird Algorithm

Zhao et al.[29] imitates the behaviour of the famous hummingbird and introduced the novel bio-inspired optimization i.e. Artificial Hummingbird Algorithm (AHA). These birds eat the insects, nectar and sweet liquid from flowers to keep them energized and activated. These birds are considered to be very smart because of their sharp memory of foraging. Because of their sharp memory, these birds visit the best location of the food source and avoid revisiting the recently sampled flowers. Also, hummingbirds have flexible joints which makes them fly in any direction at any altitude. They can change their direction of flight at any instant of time. Diagonal flight is one of the special flight postures in which hummingbirds can stay in the air and migrate anywhere. In AHA, the intelligence and flight skills of the hummingbirds have been simulated through guidance, territorial and migrating foraging.

### 3.2.1. Initialization

Taking $m$ food sources then the population of hummingbirds are randomly initialized through Eq. (8).

$$x_i = Lb + rand(0,1).(Ub - Lb), \qquad i = 1, 2, ..., m \tag{8}$$

where $Ub$ and $Lb$ are upper and lower bounds respectively. $rand(0,1)$ indicates the vector in the range [0,1], $x_i$ represents the candidate's solution. The $i^{th}$ hummingbird visits the $j^{th}$ food source as shown in Eq. (9).

$$VT_{i,j} = \begin{cases} 0, i \neq j \\ null, i \neq j \end{cases} \quad i = 1, 2, ..., m, \ j = 1, 2, ..., m \tag{9}$$

### 3.2.2. Foraging guidance

The foraging guidance of the hummingbirds is simulated based on their flight skills which include axial, diagonal and omnidirectional flight. The expression for the axial flight, diagonal flight and omnidirectional flight are shown in Eq.(10), Eq.(11) and Eq. (12) respectively.

$$D^i = \begin{cases} 1, & i = randi([1,d]) \\ 0, & else \end{cases} \quad i = 1, 2, ..., d \tag{10}$$

$$D^i = \begin{cases} 1, & i = w(ii), \ ii \in [1, K], \ w = randperm(K), \ K \in [2\lceil R\lceil(d-2)\rceil\rceil+1] \\ 0, & else \end{cases} \quad i = 1, 2, ..., d \tag{11}$$

$$D^i = 1, i = 1, 2, ..., d \tag{12}$$

where, $randi([1, d])$ and $randperm(K)$ depict the random integer, $R$ is a random number that falls between [0,1]. Through these intelligence and flight skills, these hummingbirds visit the target food sources to access the candidate food source.

$$Vg_i = x_{i,target}(t) + a \cdot D \cdot [x_i(t) - x_{i,target}(t)] \tag{13}$$

$$aN(0,1) \tag{14}$$

where, $x_i(t)$ and $x_{i,target}(t)$ represents the candidate and target candidate solutions respectively that $i^{th}$ hummingbirds visits. The variable $a$ is a regional foraging parameter that follows the normal distribution. The position of the $i^{th}$ hummingbird can be updated through Eq. (15).

$$x_i(t+1) = \begin{cases} x_i(t), & f(x_i(t)) \leq f(Vg_i(t+1)), \\ Vg_i(t+1), & f(x_i(t)) > f(Vg_i(t+1)), \end{cases} \tag{15}$$

where, $f(\cdot)$ is the fitness function. The hummingbird decides whether to choose the candidate food source obtained from Eq. (13) or the current candidate solution obtained from Eq. (15) based on the nectar supplement rate.

### 3.2.3. Territorial foraging

The hummingbirds once got the target candidate solution then it moves towards new food sources so that they explore the alternative solution in the adjacent areas through their flying skills and position. In this way, the hummingbirds extend their territory which can be expressed through Eq. (16).

$$Vt_i(t+1) = x_i(t) + b \cdot D \cdot x_i(t) \tag{16}$$

$$b \square N(0,1) \tag{17}$$

where $b$ is the regional foraging parameter that follows the standard normal distribution.

### 3.2.4. Migratory foraging

The hummingbirds migrate to other places when the food in the territorial area becomes less. While taking the migration into action, the migration factor will be prefixed in the AHA. The expression for migration and updation of table access is given in Eq. (18).

$$x_{poor} = Lb + r \cdot (Ub - Lb), \quad i = 1, 2, ..., m \tag{18}$$

where, $x_{poor}$ represents the worst candidate solution. In the AHA algorithm, apart from the migration factor, two other parameters are also predetermined that are population size and maximum iterations. But it is only the migration coefficient that tells whether to perform migration or not. The migration coefficient ($M$) is given by Eq. (19).

$$M = 2m \qquad (19)$$

## 4. Proposed Methodology

In the proposed scheme, the AHA is utilized to optimize the parameters of FMD taking novel HI i.e. *SIMI* as the fitness function given in Eq. (20).

$$\begin{cases} fitness = \min_{(K,L)} (SIMI_i) \\ \quad s.t. \quad K \in (3,8) \\ \quad \quad \quad L \in [20,50] \end{cases} \qquad (20)$$

$SIMI_i$ is the fitness function obtained for each modeIMIitness, $i = 1, 2, \ldots, N$. The proposed methodology is depicted in Fig. 1.

1) Initialize the AHA with a population of 30 and a maximum iteration of 20. Feed the raw signals into FMD at preset values of its parameters.

2) Decompose the raw signals into different modes.

3) Compute the fitness function i.e. SIMI for each mode till it reaches the maximum iteration.

4) Evaluate the global optimal fitness function and save the optimal values of mode number ($K$) and filter length ($L$).

5) The energy for each mode is computed. Based on which energy interval range has been decided which are then transformed into SVNSs to prepare training and testing pool.

6) The least SVNCE values between SVNSs of testing samples (obtained from unknown bearing conditions) and SVNSs of training samples (obtained from known bearing conditions) are used in the minimum argument principle to identify various bearing faults.

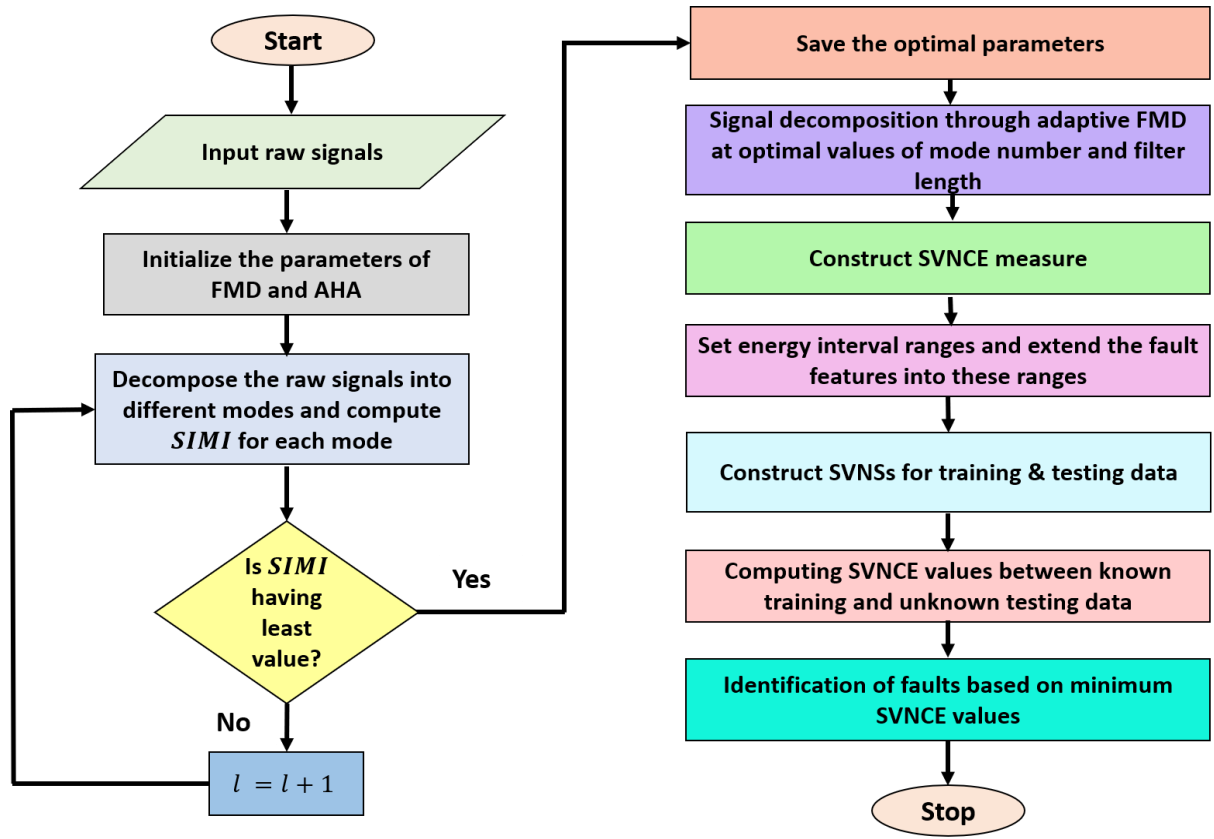

**Fig. 1.** Methodology of the proposed work

## 5. Utilising the proposed approach in real-time application

**Step 1: Data Acquisition**

The machines used in the mining industries are high power machines that have complex structure and time varying. The conveyor driving station is drived by different drives having power ranging from 630 to 1000 kW. In our case, two drives of 1000 kW power has been used. The drive unit is composed of electric motor, a coupling and two stage gearbox, that are connected with a pulley as shown in Fig. 2. The pulley is mounted on the shaft and balanced by the two set of bearings. The pulley is coated by the rubber to increase the friction between pulley and the belt. The rigid coupling has been utilised to establish the connection between pulley and gearbox.

The vibration signals has been acquired from the pulley to analyse the condtion of the bearings as shown in the Fig. 2 (c). The accelerometer is mounted on the bearing in the horizontal direction as shown in Fig. 2 (d) through screw. The defect frequencies based on rotational speed was found to be $f_{FTF} = 0.51$ Hz, $f_{BSF} = 4.45$ Hz, $f_{BFF} = 8.90$ Hz, $f_{BPFO} = $

12.34 Hz and $f_{BPFI} = 16.06$ Hz. These signals have been acquired for 2.5 seconds at sampling frequency of 19200 Hz.

The vibration signals from two cases of bearing i.e. defect free and defected bearing have been acquired as shown in Fig. 3. The vibration from the gearbox has also been noticed in the vibration signal of the bearing which almost covers the signal of interest.

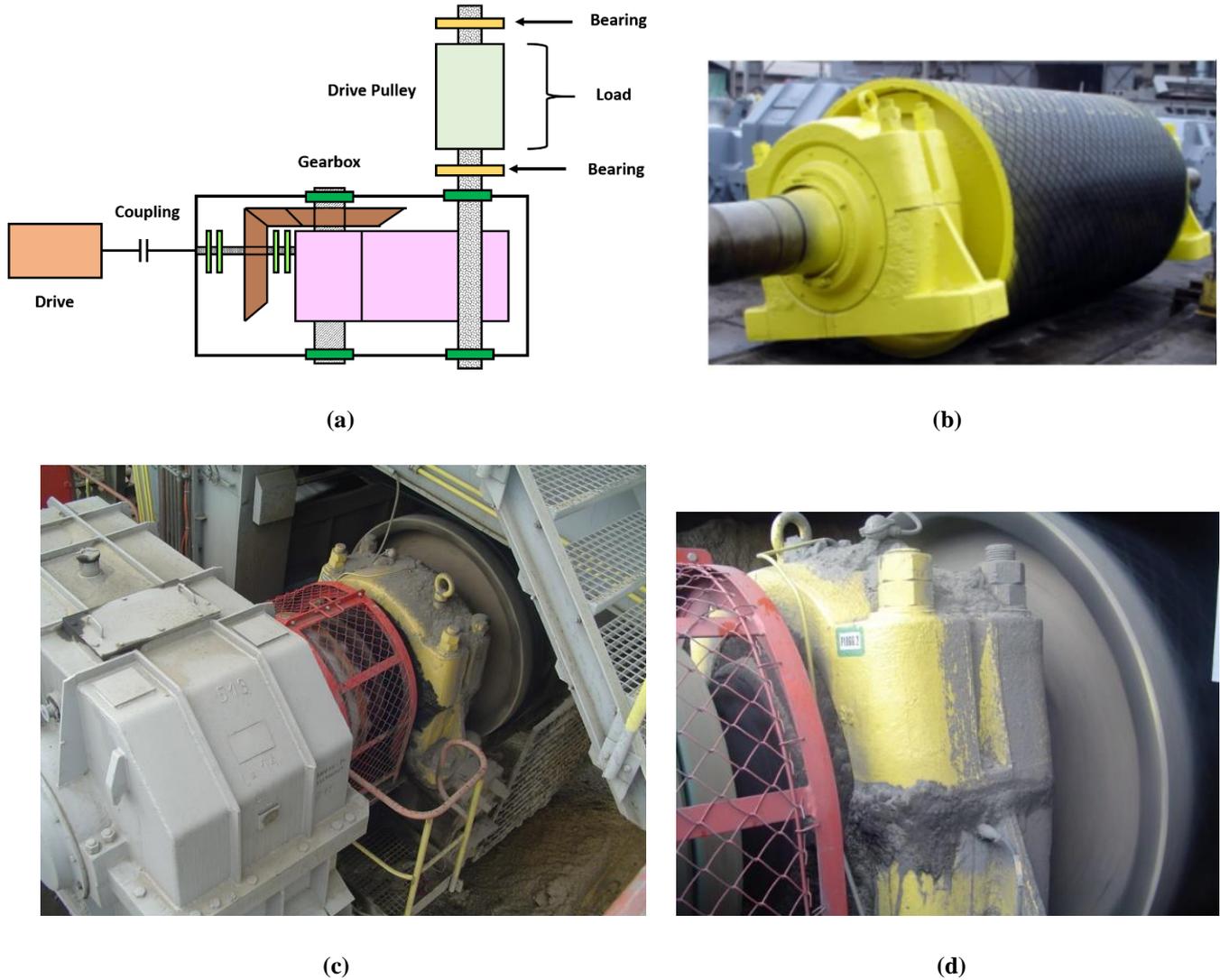

**Fig. 2.** Test rig **(a)** Graphic representation **(b)** pulley with bearing housing mounted on shaft **(c)** view on joint of output shaft in gearbox with pulley, and **(d)** view on sensor location on pulley [30][31]

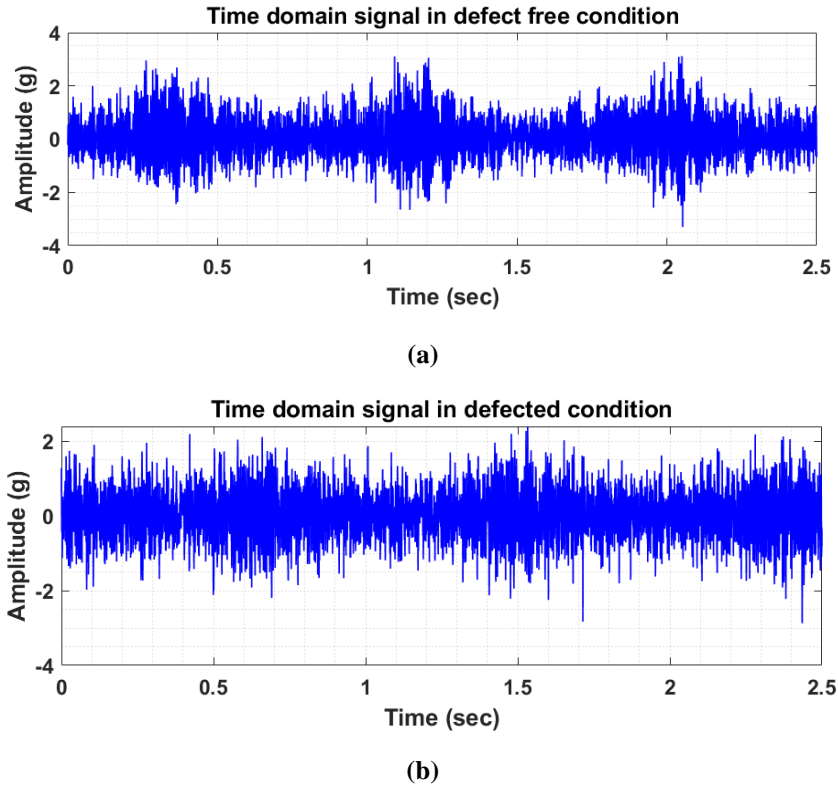

**Fig. 3.** Raw signals under **(a)** healthy condition **(b)** unhealthy condition

It is obvious that the fault impulses are obtained by random components, making it impossible to detect the fault directly from the raw signal.

**Step 2: Decomposition of raw signals based on FMD**

The raw signals are disintegrated through FMD to identify the bearing defects. The FMD's parameters i.e., mode number ($K$) and filter length ($L$) have been evaluated through an optimization algorithm (AHA) based on the minimum value of developed *SIMI* index. $K$ and $L$ in the case of healthy conditions are found to be 7 and 30. The optimal values of the same parameters for defected conditions have been obtained as 7 and 42. The different modes from different health conditions are shown in Fig. 4. Further four prominent modes has been selected from each health condition of analysis based on the minimum value of *SIMI* index. The values for *SIMI* index obtained from healthy bearing are 0.0729, 0.1868, 0.3810, 0.4090, 0.4645, 0.3994 and 0.3015. Whereas for the unhealthy bearing the values of *SIMI* index are 0.0798, 0.0139, 0.0112, 0.0198, 0.0157, 0.0120 and 0.2494.

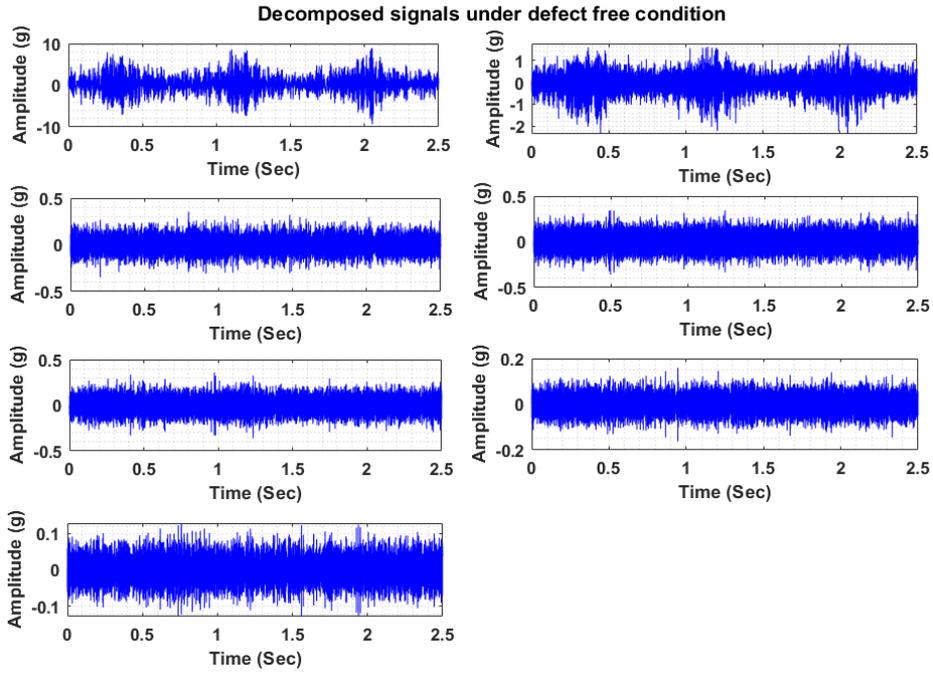

(a)

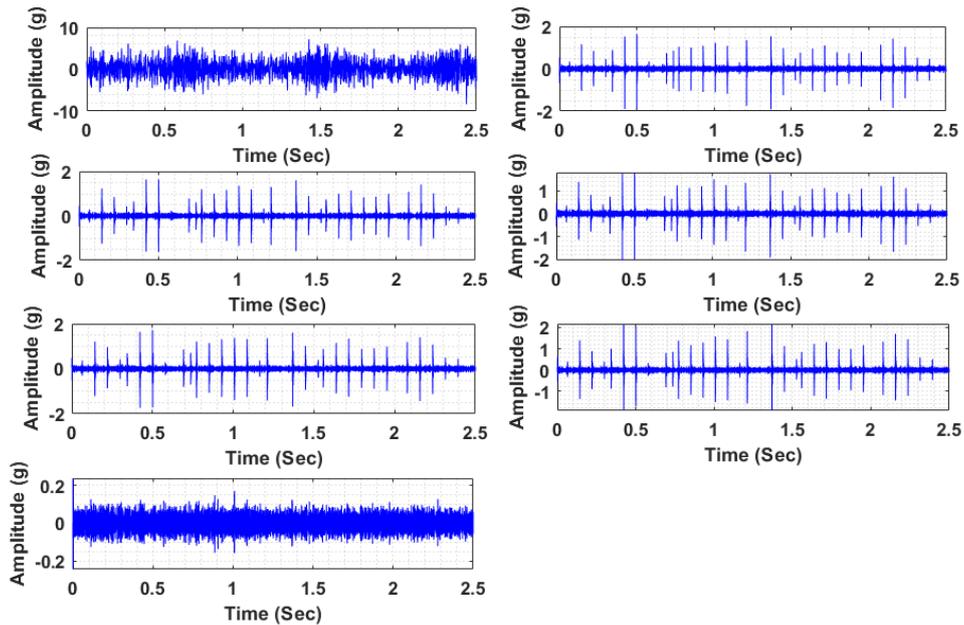

(b)

**Fig. 4.** Decomposed signals obtained from **(a)** Healthy condition **(b)** Inner race defect **(c)** Outer race defect **(d)** Roller defect

**Step 3: Computation of energy eigen vector**

The FMD is used to decompose raw signals at optimal mode number and filter length which are further utilized to extract the faulty information from the bearing. The different faulty conditions are depicted by the set $D = (D_1, D_2)$, where $D_1$ = healthy bearing, $D_2$ = defected bearing. The energy eigenvector is constructed utilizing the energy interval which is depicted through $E = [E^1, E^2, E^3, E^4]$. While carrying out this analysis, the weight $w_i = \frac{1}{2}(i = 1,2)$ is assigned to different health conditions. These eigenvalues are normalized through Eq. (21).

$$E_{norm}^k = \frac{E^k - \min(E^k)}{\max(E^k) - \min(E^k)}, \quad k = 1,2,3,4 \tag{21}$$

where $max(E^k)$ and $min(E^k)$ are the maximum and minimum values of $k^{th}$ mode, respectively.

**Step 4: Defining the fault features energy interval based on lower and upper bounds**

For each sub-frequency band, the distribution of energy for modes produced by FMD varies depending on the bearing's unknown state of health. The obtained data is used to define the lower and upper bound which are further utilised to evaluate the energy interval range. The results obtained are tabulated in Table 1.

**Table 1**
Establishing the energy interval ranges for $ID_K(K = 1,2)$.

| Fault Type | Defining the lower and upper energy at each sub-frequency band | SVNSs form of energy interval range |
|---|---|---|
| $D_1$ | [0.01 0.9968] [0.01 0.9991] [0.01 0.9778] [0.01 0.9980] | [0.01 0.9868 0.0032] [0.01 0.9891 0.0009] [0.01 0.9678 0.0222] [0.01 0.9880 0.0020] |
| $D_2$ | [0.01 0.9710] [0.01 0.9727] [0.01 0.9946] [0.01 0.9624] | [0.01 0.9610 0.0290] [0.01 0.9627 0.0273] [0.01 0.9846 0.0054] [0.01 0.9524 0.0376] |

**Step 5: Preparation of training and testing samples using SVNSs**

In this step, SVNSs are computed based on the energy interval range for modes obtained from different health conditions. Supposing the lower and upper bounds of the $i^{th}$ energy eigenvalue for $D_K(K = 1,2,3,4)$ are indicated by $LB_{D_K}(x_i)$ and $UB_{D_K}(x_i), (K = 1,2,3,4)$ $(i = 1,2,3,4)$ respectively. Then

$$D_K = \{< x_1, [LB_{D_K}(x_1), UB_{D_K}(x_1)] >, < x_2, [LB_{D_K}(x_2), UB_{D_K}(x_2)] >, <$$
$$x_3, [LB_{D_K}(x_3), UB_{D_K}(x_3)] >, < x_4, [LB_{D_K}(x_4), UB_{D_K}(x_4)] >\} \tag{22}$$

Let $f_{D_K}(x_i) = 1 - UB_{D_K}(x_i)$ and $i_{D_K}(x_i) = 1 - f_{D_K}(x_i) - LB_{D_K}(x_i)$. A constraint is put on the values of $i_{D_K}(x_i)$ is 0.001 which means if the obtained value is less than 0.01 then to satisfy the non-negative condition: $0 \leq LB_{D_K}(x_i) + i_{D_K}(x_i) + f_{D_K}(x_i) \leq 3$ where $LB_{D_K}(x_i), i_{D_K}(x_i), f_{D_K}(x_i): X \rightarrow [0,1]$. Further, the set $D_K$ is represented in SVNSs which are shown in Table 2. The modified form is shown in Eq. (22).

$$D_K = \{< x_1, [LB_{D_K}(x_1), i_{D_K}(x_1), f_{D_K}(x_1)] >, < x_2, [LB_{D_K}(x_2), i_{D_K}(x_2), f_{D_K}(x_2)] >, <$$
$$x_3, [LB_{D_K}(x_3), i_{D_K}(x_3), f_{D_K}(x_3)] >, < x_4, [LB_{D_K}(x_4), i_{D_K}(x_4), f_{D_K}(x_4)] >\} \quad (23)$$

Similarly, the same procedure is implemented for testing samples $T_{T_j}(j = 1,2,3,4)$ and shown as:

$$T_{T_1} = \{< x_1, [\ 0.01\ \ 0.9856\ 0.0044] >, < x_2, [\ 0.01\ \ 0.9838\ 0.0062] >, <$$
$$x_3, [0.01\ \ 0.9801\ 0.0099] >, < x_4, [\ 0.01\ \ 0.9462\ 0.0438] >\} \quad (24)$$

$$T_{T_2} = \{< x_1, [\ 0.01\ \ 0.9096\ 0.0804] >, < x_2, [\ 0.01\ \ 0.8765\ 0.1135] >, <$$
$$x_3, [0.01\ \ 0.8998\ 0.0946] >, < x_4, [\ 0.01\ \ 0.8998\ 0.0902] >\} \quad (25)$$

The obtained SVNSs are used to reinterpret the Theorems given in Appendix by updating $\mu_A(x), i_A(x),$ and $f_A(x)$ with $LB_{D_K}(x), i_{D_K}(x),$ and $f_{D_K}(x)$ respectively, the weighted SVNCE measure is represented as:

$$T(T_{T_j}, D_K)(j = 1,2,3,4; K = 1,2,3,4) = \sum_{i=1}^{4} w_i \left[ \left(2 + LB_{D_K}(x_i) + LB_{T_{T_j}}(x_i)\right) log_2 \left[\frac{2 + LB_{D_K}(x_i) + LB_{T_{T_j}}(x_i)}{4 + LB_{D_K}(x_i) + LB_{T_{T_j}}(x_i) + 2\sqrt{LB_{D_K}(x_i) \times LB_{T_{T_j}}(x_i)}}\right] + \left(4 - LB_{D_K}(x_i) - LB_{T_{T_j}}(x_i)\right) log_2 \left[\frac{4 - LB_{D_K}(x_i) - LB_{T_{T_j}}(x_i)}{6 - LB_{D_K}(x_i) - LB_{T_{T_j}}(x_i) + 2\sqrt{\left(1 - LB_{D_K}(x_i)\right) \times \left(1 - LB_{T_{T_j}}(x_i)\right)}}\right]\right] + \sum_{i=1}^{4} w_i \left[\left(2 + i_{D_K}(x_i) + i_{T_{T_j}}(x_i)\right) log_2 \left[\frac{2 + i_{D_K}(x_i) + i_{T_{T_j}}(x_i)}{4 + i_{D_K}(x_i) + i_{T_{T_j}}(x_i) + 2\sqrt{i_{D_K}(x_i) \times i_{T_{T_j}}(x_i)}}\right] + \left(4 - i_{D_K}(x_i) - i_{T_{T_j}}(x_i)\right) log_2 \left[\frac{4 - i_{D_K}(x_i) - i_{T_{T_j}}(x_i)}{6 - i_{D_K}(x_i) - i_{T_{T_j}}(x_i) + 2\sqrt{\left(1 - i_{D_K}(x_i)\right) \times \left(1 - i_{T_{T_j}}(x_i)\right)}}\right]\right] + \sum_{i=1}^{4} w_i \left[\left(2 + f_{D_K}(x_i) + \right.\right.$$

$$f_{T_{T_j}}(x_i)\bigg) log_2 \left[ \frac{2+f_{D_K}(x_i)+f_{T_{T_j}}(x_i)}{4+f_{D_K}(x_i)+f_{T_{T_j}}(x_i)+2\sqrt{f_{D_K}(x_i)\times f_{T_{T_j}}(x_i)}} \right] + \bigg( 4 - f_{D_K}(x_i) -$$

$$f_{T_{T_j}}(x_i)\bigg) log_2 \left[ \frac{4-f_{D_K}(x_i)-f_{T_{T_j}}(x_i)}{6-f_{D_K}(x_i)-f_{T_{T_j}}(x_i)+2\sqrt{\left(1-f_{D_K}(x_i)\right)\times\left(1-f_{T_{T_j}}(x_i)\right)}} \right] \Bigg] \quad (26)$$

**Step 6: Computing the SVNCE values between SVNSs of training and testing samples**

The SVNCE values between SVNSs of the testing sample and training samples is evaluated and shown in Table 2.

**Table 2**
Recognition of bearing defects through proposed method

| Measure | SVNCE Values | | Fault diagnosis order | Fault condition recognized | Authentic fault condition |
|---|---|---|---|---|---|
| | $ID_1$ | $ID_2$ | | | |
| $FMD + SVNCE$ | 0.0091 | 0.0126 | $ID_2 > ID_1$ | Healthy bearing | Healthy bearing |
| $FMD + SVNCE$ | 0.0784 | 0.0406 | $ID_1 > ID_2$ | Defected bearing | Defected bearing |

**Step 7: Recognition of bearing defects using minimum SVNCE values**

Using the minimum argument principle, the minimum value was determined through $T_{T_1}$ and $D_K(K = 1,2)$ is **0.0091** that matches the healthy condition of the bearing $(D_1)$. The minimum value of **0.0406** is computed between $T_{T_2}$ and $D_K(K = 1,2)$ which represents the unhealthy condition of the bearing $(D_2)$. The procedure is repeated for 30 cases of each health condition. The performance of the proposed method is tabulated in Table 3. The proposed method has also been compared with the existing methods such as EEMD and VMD whose results are tabulated in Table 4.

**Table 3**
Performance of the proposed technique for bearimg defect recognition

| | | Accuracy (%) | | Overall Accuracy (%) |
|---|---|---|---|---|
| | | Healthy bearing $(D_1)$ | Defected bearing $(D_2)$ | |
| $FMD + SVNCE$ | Healthy bearing | 100 | 0 | 99.6 |
| | Defected bearing | 0 | 99.2 | |

**Table 4**
Performance of different decomposition techniques for fault identification

|  | Accuracy (%) | | Overall Accuracy (%) |
| --- | --- | --- | --- |
|  | Defect free ($D_1$) | Defected bearing ($D_2$) |  |
| **Existing method based on EEMD** | 78.5 | 83.6 | 81.05 |
| **Existing method based on VMD** | 77.9 | 81.4 | 79.65 |

## 6. Conclusion

In this research, a new scheme based on extraction SVNCE from modes obtained from adaptive FMD has been developed to identify the bearing defects.

- In order to extract more explicit and incipient fault characteristics, an adaptive FMD was proposed through optimal selection of mode number and filter length.
- A novel health indicator i.e. *SIMI* has been developed which acts as the fitness function during the optimization process. *SIMI* analyses the fault characteristic information which characterizes the signal's periodic impulse as well as its degree of resistance to impulsive noises.
- The energy of modes obtained by adavptive FMD sets the interval range based on energy eigenvalues. Further, the SVNSs are constructed utilizing the energy interval range. The bearing defects are automatically identified using minimum values of SVNCE obtained between SVNSs of training samples (obtained from bearing conditions) and SVNSs of testing samples (obtained from unknown bearing conditions).
- The proposed method is superior and robust to other decomposition techniques as it gives overall accuracy of 99.6 % which much higher than other existing methods.

**Acknowledgments**

The work is supported by the National Center of Science, Poland under Sheng2 project No. UMO-2021/40/Q/ST8/00024 "NonGauMech - New methods of processing non-stationary signals (identification, segmentation, extraction, modeling) with non-Gaussian characteristics for the purpose of monitoring complex mechanical structures".